\documentclass[12pt]{iopart}

\usepackage{bm}
\usepackage{epsf}
\usepackage{subfigure} 
\usepackage[english]{babel}
\usepackage{dcolumn}
\usepackage{graphicx}

\begin{document}
\bibliographystyle{unsrt}


\title{Photodissociation of the HeH$^{\bm +}$ molecular ion}

\author{Irina Dumitriu and Alejandro Saenz}
\address{Humboldt-Universit\"at zu Berlin, Institut f\"ur Physik, 
AG Moderne Optik, Hausvogteiplatz 5-7, D-10117 Berlin, Germany.}
\ead{\mailto{Alejandro.Saenz@physik.hu-berlin.de}}

\date{\today}

\begin{abstract}
The photodissociation cross section of the molecular ion HeH$^+$ was 
calculated within the Born-Oppenheimer approximation for a parallel, a 
perpendicular, and an isotropic orientation of the molecular axis with respect 
to the field, considering also different initial vibrational and rotational 
states. The results were compared to recent data from a free-electron laser 
experiment performed at the FLASH facility 
[H.B.~Pedersen {\it et al.}, Phys.\,Rev.\,Lett.\,98, 223202, (2007)].   
Within the experimental uncertainties theoretical and 
experimental results are compatible with each other. 
\end{abstract}

\pacs{31.50.Gh, 33.80.Gj}
\submitto{\JPB}
\maketitle


\section{Introduction}
\label{intro}
The helium-hydride ion HeH$^{\bm +}$ is the simplest heteronuclear 
two-electron 
system and the only well bound molecular system of hydrogen and helium, 
the most abundant elements in the universe. This motivated many calculations
and 
experiments performed on this molecular ion which is of interest for the 
chemistry of astrophysical objects, for the tritium neutrino mass experiments, 
or simply by itself as a model system (see, e.\,g., 
\cite{dia:mich66,dia:gree74a,dia:gree74b,dia:gree76,dia:gree78,dia:saha78,dia:basu84,dia:saen03,dia:pava05,dia:zhou06}  
and references therein). 
More recently, HeH$^{\bm +}$ is drawing special attention due to 
an experiment performed at the free-electron laser FLASH \cite{dia:pede07} 
where the absolute photodissociation cross section at a photon 
energy of 38.7 eV was measured. 

The photoabsorption spectrum of HeH$^{\bm +}$ was previously 
investigated theoretically, together with photoionization, only for 
the parallel orientation of the molecular axis with 
respect to the field \cite{dia:saen03}. 
A more recent study on the photoionization of HeH$^{\bm +}$ \cite{dia:fern07} 
treated also the perpendicular orientation, but 
for the photodissociation into electronic bound states, to the authors' 
knowledge, the only work 
considering both the parallel and the perpendicular orientation is still 
the one of Basu and Barua from 1984 \cite{dia:basu84}. There, however, the $\Pi$ 
contribution taken into account is only the one of the lowest state, 
while the FLASH experiment shows an important contribution of higher lying 
$\Pi$ states \cite{dia:pede07}. 

Motivated by the FLASH experiment, the photodissociation cross section of 
HeH$^{\bm +}$ has been 
calculated for parallel, perpendicular, as well as isotropic orientations 
of the molecular axis with respect to the field, 
for transitions starting from the lowest lying vibrational state and the 
first or second rotational level, as well as for a mixture of 
initial vibrational states. Since the experiment measures only the
dissociation into He + H$^{\bm +}$, a comparison to the experimental 
data required resolving different dissociation channels. The influence 
of non-adiabatic effects on the channel-resolved results is discussed. 
The paper is organized as follows: 
section \ref{method} presents the method and the 
computational details, section \ref{results} discusses the  
results and a summary is given in Section \ref{summ}. 
%
%
\section{Method and computational details}
\label{method}
In atomic units ($1\,E_h = 27.21139\,$eV and $1\,a_0 =
5.29177\,\cdot\,10^{-11}$\,m) and the non-relativistic dipole approximation 
the photodissociation cross section is given according to 
Fermi's Golden Rule by%
\begin{equation}
\label{eq:cross}
   \sigma (\omega) = g \;\frac {4 \pi^2} {c} \omega \; 
                     |\langle \Psi_{n' \varepsilon' J' m_{J}'} | \hat{\rm D} | 
                      \Psi_{n \nu J m_J} \rangle|^2 \cdot \delta(\omega - 
                      [E_{n' \varepsilon' J' m_{J}'} - E_{n \nu J m_J}]) 
\end{equation}
where $\omega$ is the photon energy and $\hat{\rm D}$ is the 
dipole operator in length gauge. For a diatomic molecule, it is sufficient 
to calculate the single-photon absorption cross section for a 
parallel and a perpendicular orientation of the molecular axis 
with respect to the electric field component. The partial parallel 
(perpendicular) cross section for an isotropic molecular ensemble 
is then obtained by a multiplication of the cross section for a molecule 
oriented parallel (perpendicular) to the field with the statistical 
factor $g=1/3$ ($g=2/3$). The total cross section for an isotropic 
molecular ensemble is simply the sum of these two partial 
cross sections (see also \cite{dia:saen03,aies:dumi07} for more details 
on $g$). 

In equation (\ref{eq:cross}), $\Psi_{n \nu J m_J}$ and $\Psi_{n'\varepsilon' J' m_{J}'}$ are 
the rovibronic wavefunctions of the initial and final state, with energies
$E_{n \nu J m_J}$ and $E_{n'\varepsilon' J' m_{J}'}$ respectively. 
Within the Born-Oppenheimer (BO) approximation the former is given as%
\begin{equation}
\label{eq:wavef}
\Psi_{n \nu J m_J} (\vec{r}_1,\vec{r}_2,\vec{R}\,) = 
          \psi_{n}(\vec{r}_1,\vec{r}_2; R) \; 
               \frac{\Phi_{\nu J}^{(n)}(R)}{R}\; Y_{J m_J}(\Omega)
\end{equation}
where $\psi_{n}(\vec{r}_1,\vec{r}_2; R)$ is the 
electronic, $\Phi_{\nu J}^{(n)}(R)/R$ the vibrational, and the spherical 
harmonic $Y_{J m_J}$ the rotational wavefunction. The index $n$ specifies 
the electronic state (BO potential curve) including its 
symmetry (the only relevant ones in this work are $^1\Sigma$ and $^1\Pi$), 
while $\nu$ and $\{J,m_J\}$ are the vibrational and rotational quantum numbers, 
respectively. The final-state wavefunction can be written in an analogous way, 
but in the here studied case of photodissociation $\varepsilon'$ is a
continuous index, while $n$, $n'$, $\nu$, $J$, and $J'$ are discrete ones. 
Correspondingly, $\psi_{n}$, $\psi_{n'}$, $\Phi_{\nu J}^{(n)}$, and, of course, 
$Y_{J m_J}$ as well as $Y_{J' m_{J}'}$ are normalized to unity, while 
$\Phi_{\varepsilon' J'}^{(n')}$ is energy normalized, 
$\langle \Phi_{\varepsilon' J'}^{(n')} | \Phi_{\varepsilon'' J'}^{(n')} \rangle = 
\delta(\varepsilon'-\varepsilon'')$.

The transition amplitudes in equation\,(\ref{eq:cross}) are obtained by a
calculation of the integral%
\begin{equation} 
\label{eq:dip}
   \langle \Psi_{n'\varepsilon' J' m_{J}'} | 
            \hat{\rm D} | \Psi_{n \nu J m_J} \rangle \; = \; \int \,
                     \Phi_{\varepsilon' J'}^{*\,(n')}(R) \,D_{n' n}(R) \, 
                                                  \Phi_{\nu J}^{(n)}(R)\; {\rm d} R
\end{equation} 
where  $D_{n' n}(R)$ denotes the electronic transition dipole 
moment as a function of the internuclear distance. It is defined by
\begin{equation}
\label{eq:dipr}
  D_{n' n}(R) = \int \psi^*_{n'}(\vec{r}_1,\vec{r}_2; R)\; 
            \hat{d}\;\psi_{n}(\vec{r}_1,\vec{r}_2; R) \; {\rm d}^3 r 
\end{equation}
where $\hat{d}$ is the electronic dipole operator in length gauge,
$\hat{d}=\vec{\epsilon} \,\cdot\, (\vec{r}_1 + \vec{r}_2)$ with 
$\vec{\epsilon}$ being the polarization vector. (The contribution 
of the nuclear dipole operator vanishes, since in the present work only 
electronic transitions with $n\neq n'$ are considered.) 

The electronic wavefunctions and transition dipole moments are calculated 
using the method described in \cite{dia:vann04} which is thus only briefly 
sketched in the following. 
First, the one-electron Schr\"odinger equation (OESE) is solved 
in an elliptical ``box'' within the fixed-nuclei approximation, using a 
{\it B}-spline basis set and the prolate-spheroidal coordinates 
($1 \leq \xi < \infty , \, -1 \leq \eta \leq 1 , \, 0 \leq \phi < 2\pi$).  
The box boundary is defined by $\xi_{\rm max}$. The two-electron 
Schr\"odinger equation is solved using the configuration-interaction (CI) 
method in which the configurations are expanded in terms of the 
one-electron orbitals obtained from the solution of the OESE. This yields 
the BO potential curves and the two-electron wavefunctions 
$\psi$. The latter are used to evaluate the electronic transition dipole 
moments $D_{n' n}$. All the results shown in this work 
were calculated in the length gauge, but agreement with velocity gauge 
was also investigated.

The nuclear motion is solved in the adiabatic BO potential curves by expanding 
the nuclear wavefunctions in {\it B} splines for the radial part times 
spherical harmonics describing the angular part. 
The radial Schr\"odinger equation is solved in a spherical ``box'' whose 
boundary is defined by R$_{\rm max}$. The box leads to a discretization of the
vibrational continua. The resulting vibrational wavefunctions 
$\Phi_{\varepsilon' J}^{(n')}$ are therefore all discrete and normalized 
to unity. For the dissociative continuum the renormalization factors that 
lead to an energy normalization can simply be obtained from the density 
of states. 

The present calculations were directly motivated by the recent FEL experiment 
\cite{dia:pede07}. Therefore, the basis sets were chosen for the purpose of 
computing photodissociation cross sections that take into account the 
contribution of as many Rydberg states as possible (the experimental results 
indicate the importance of highly excited states). In the present calculation 
35 $\Sigma$ and 25 $\Pi$ electronic bound states were finally included.  
A physical box ($r_{\rm max} = \frac{R}{2}\xi_{\rm max}$) equal to $40\,a_0$ was 
used, changing the size of the elliptical box $\xi_{\rm max}$ synchronously 
with the variation of the internuclear distance $R$, 40 {\it B} splines 
of order 8 were used for the $\xi$ direction and 20 {\it B} splines 
of order 6 for the $\eta$ direction. The BO potential curves were calculated 
in an $R$ interval in between 
0.1 and $60\,a_0$. The latter limit was mainly needed for a determination of 
the dissociation channel, i.\,e.\ to find out into which atomic or ionic 
fragments HeH$^+$ dissociates. The basis set used for calculating the 
vibrational wavefunctions $\Phi$ comprised a box R$_{\rm max} = 20\,a_0$ 
and 900 {\it B} splines of order 10.

%
\section{Results}
\label{results}

\begin{figure}[th]
\epsfxsize 0.75\hsize
\centerline{\mbox{\epsfbox{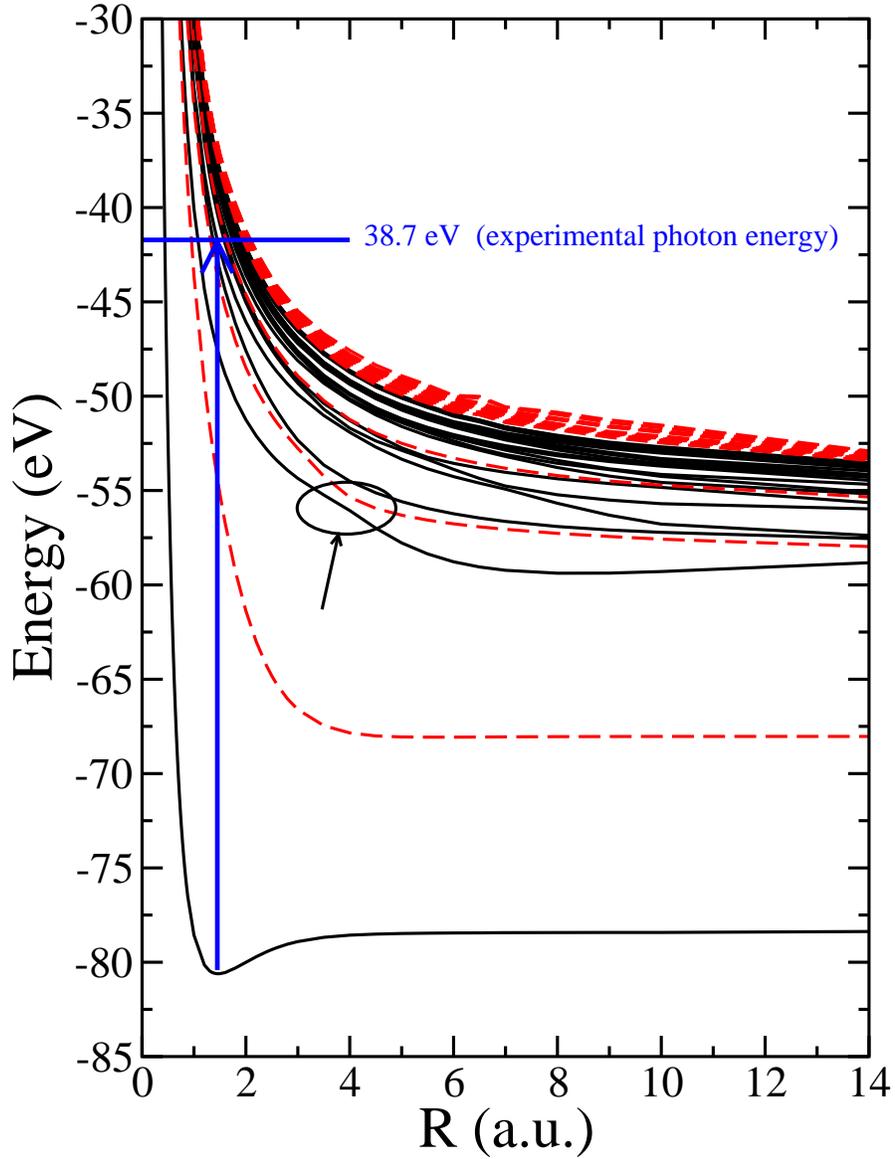}}}
\caption{The potential curves of the $^1\Sigma$ states of HeH$^{\bm +}$ 
(included in the present calculation): 
the full black lines represent states dissociating adiabatically into  
He + H$^{\bm +}$, while the red dashes represent 
the H + He$^{\bm +}$ channel. The black arrow points at an (also encircled) 
avoided crossing that is expected to influence the channel-resolved 
results. The blue vertical arrow marks the experimental photon energy used in 
\cite{dia:pede07}.}
\label{potsig}
\end{figure}

\begin{figure}[th]
\epsfxsize 0.7\vsize
\epsfxsize 0.75\hsize
\centerline{\mbox{\epsfbox{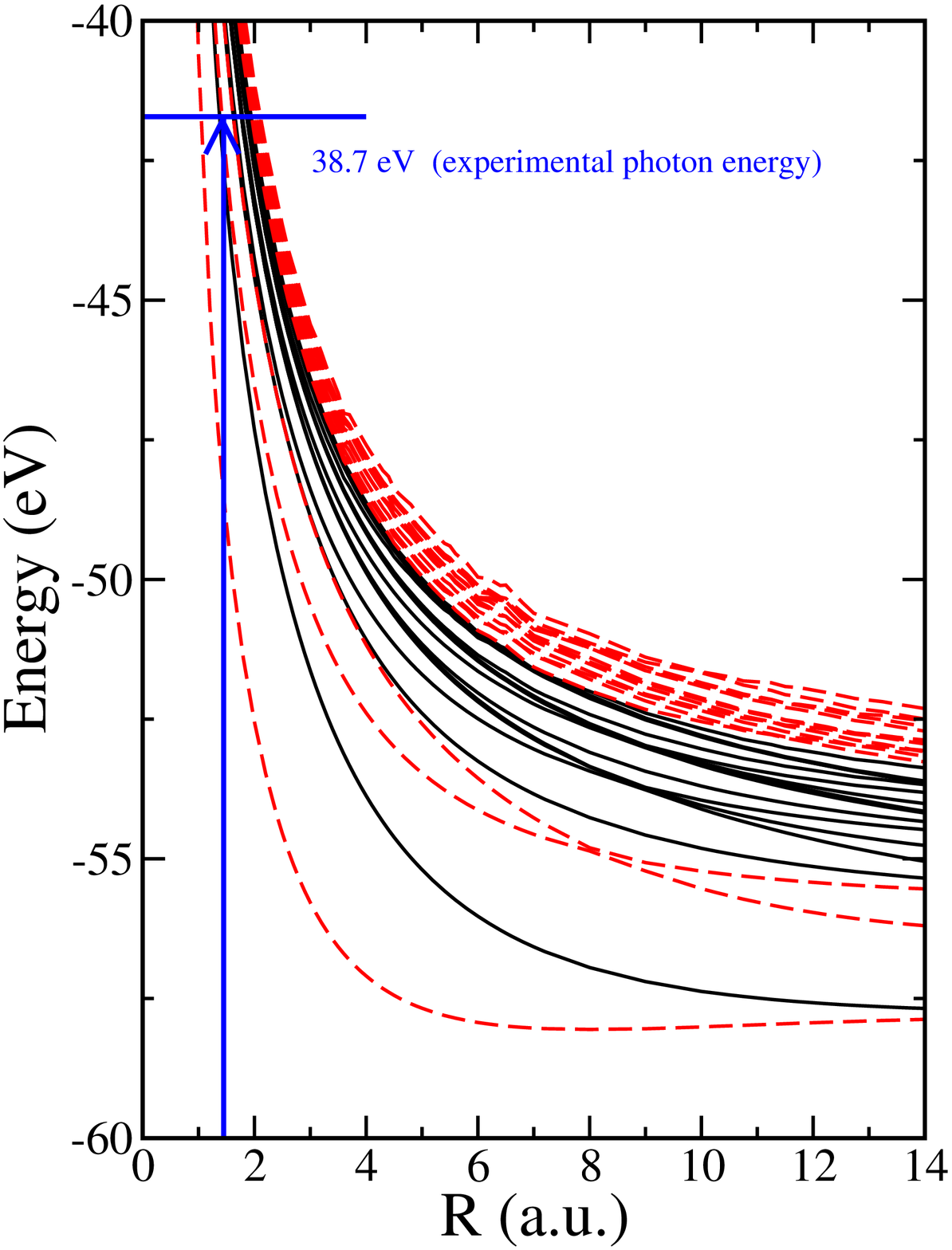}}}
\caption{As figure \ref{potsig}, but the $^1\Pi$ states are shown.}
\label{potpi}
\end{figure}

Calculations were performed for transitions starting from the lowest lying 
vibrational state of the electronic ground state to all the calculated 
electronic bound states, for both parallel and perpendicular 
orientation of the molecular axis with respect to the field. The results 
shown here are for transitions starting in the rotational level $J = 1$, which 
is the most probable value of $J$ at room temperature \cite{dia:basu84}, and 
going to $J' = 0, 2$ for the parallel orientation (according to the 
selection rule $\Delta J = \pm 1$, $\Delta J \neq 0$) and to $J' = 0, 1, 2$
for 
the perpendicular transitions (selection rule: $\Delta$J = 0, $\pm$ 1). Since 
previous calculations \cite{dia:saen03} were performed for transitions starting 
from $J = 0$, the spectrum for these transitions is also shown for 
comparison, but the 
final result is not significantly influenced by the rotational motion within 
the considered $J$ values.

\subsection{Energies, cross section, and comparison to previous results}

Calculations of electronic energies and dipole moments for different values 
of the internuclear separation were performed. For checking the 
accuracy of these calculations, results were compared to literature 
(\cite{dia:gree74a,dia:gree74b,dia:saen03}) and good 
agreement was found for the basis sets described in Section \ref{method}. 
Plots of the computed potential curves for the $^1\Sigma$ and $^1\Pi$ 
symmetries are shown in figures \ref{potsig} and \ref{potpi}, respectively. 
These potential curves were used in the subsequent calculation of the 
photodissociation cross sections. Clearly, all potential curves are 
purely dissociative within the Franck-Condon window of the electronic 
ground state. One-photon transitions from the ground to the excited 
states lead therefore to dissociation. 

\begin{figure}[th]
\epsfxsize 1.00\hsize
\centerline{\mbox{\epsfbox{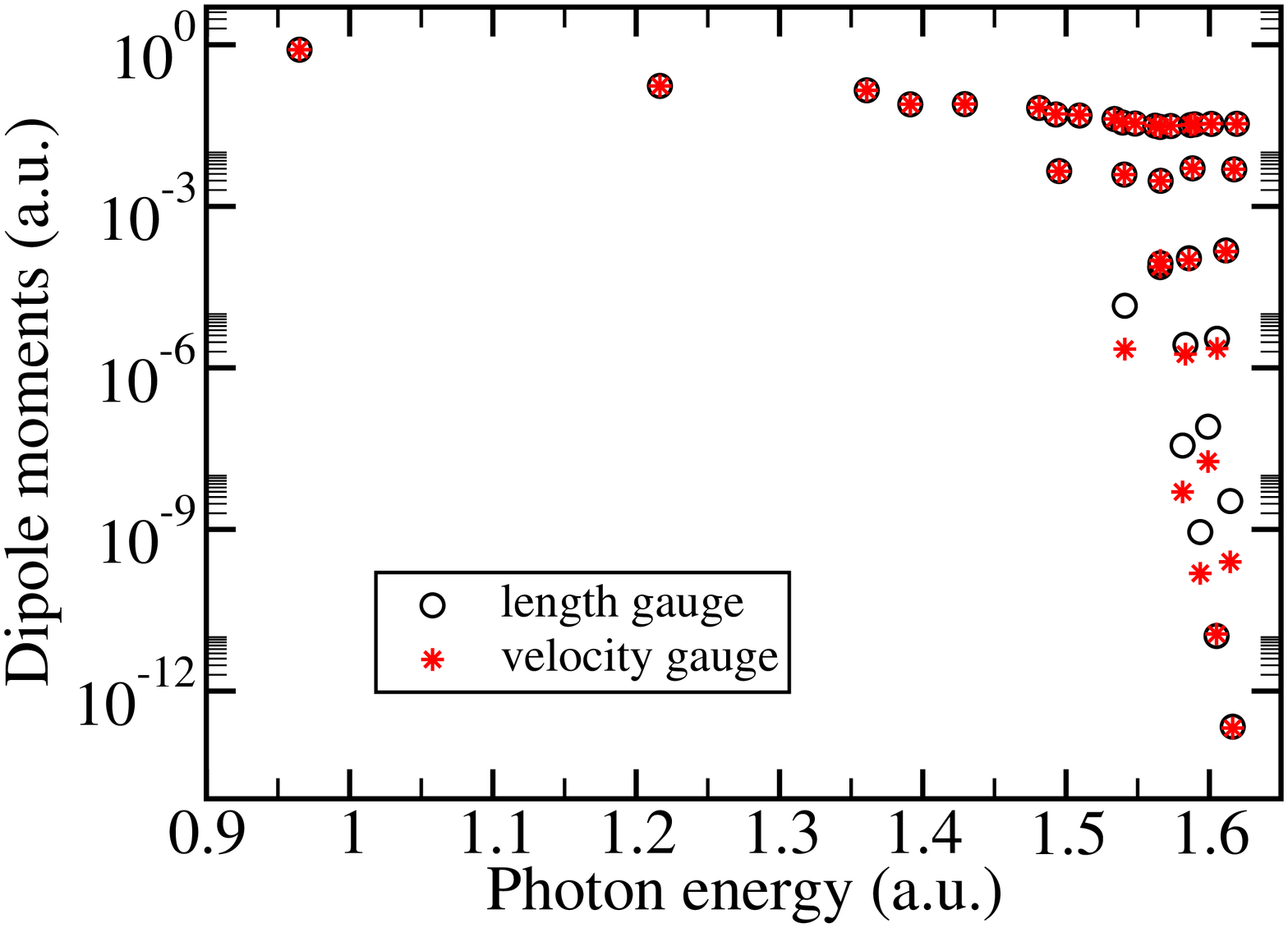}}}
\caption{The absolute value of electronic transition dipole moments for the
         equilibrium internuclear distance $R = 1.45\,a_0$, parallel
         orientation. The black circles are the length gauge results, while
         the red crosses are the velocity gauge dipole moments.} 
\label{fig:dipoles}
\end{figure}

The gauge dependence of the results was investigated by comparing 
the electronic transition dipole moments obtained in either the length 
or the velocity gauge. The latter moments are obtained from equation 
(\ref{eq:dipr}), if the dipole operator is expressed as 
$\hat{d}_v=\vec{\epsilon} \,\cdot\, (\vec{\nabla}_1 + \vec{\nabla}_2)$ and
multiplied by $\frac{1}{\omega}$ ($\omega$ being the photon energy) for
comparing to the length gauge results. 
Figure \ref{fig:dipoles} shows a comparison of the transition 
dipole moments obtained in the two gauges for the equilibrium 
internuclear separation $R_{\rm eq}= 1.45\,a_0$ and for a parallel orientation 
(transitions to $^1\Sigma$ states). The overall agreement is 
evidently very good. Noticeable deviations are only found for 
some very high lying electronic Rydberg states that possess, 
however, a very small transition moment. These deviations do 
therefore have an extremely small influence on the cross-sections 
obtained from all electronic states.

\begin{figure}[th]
\epsfxsize 1.00\hsize
\centerline{\mbox{\epsfbox{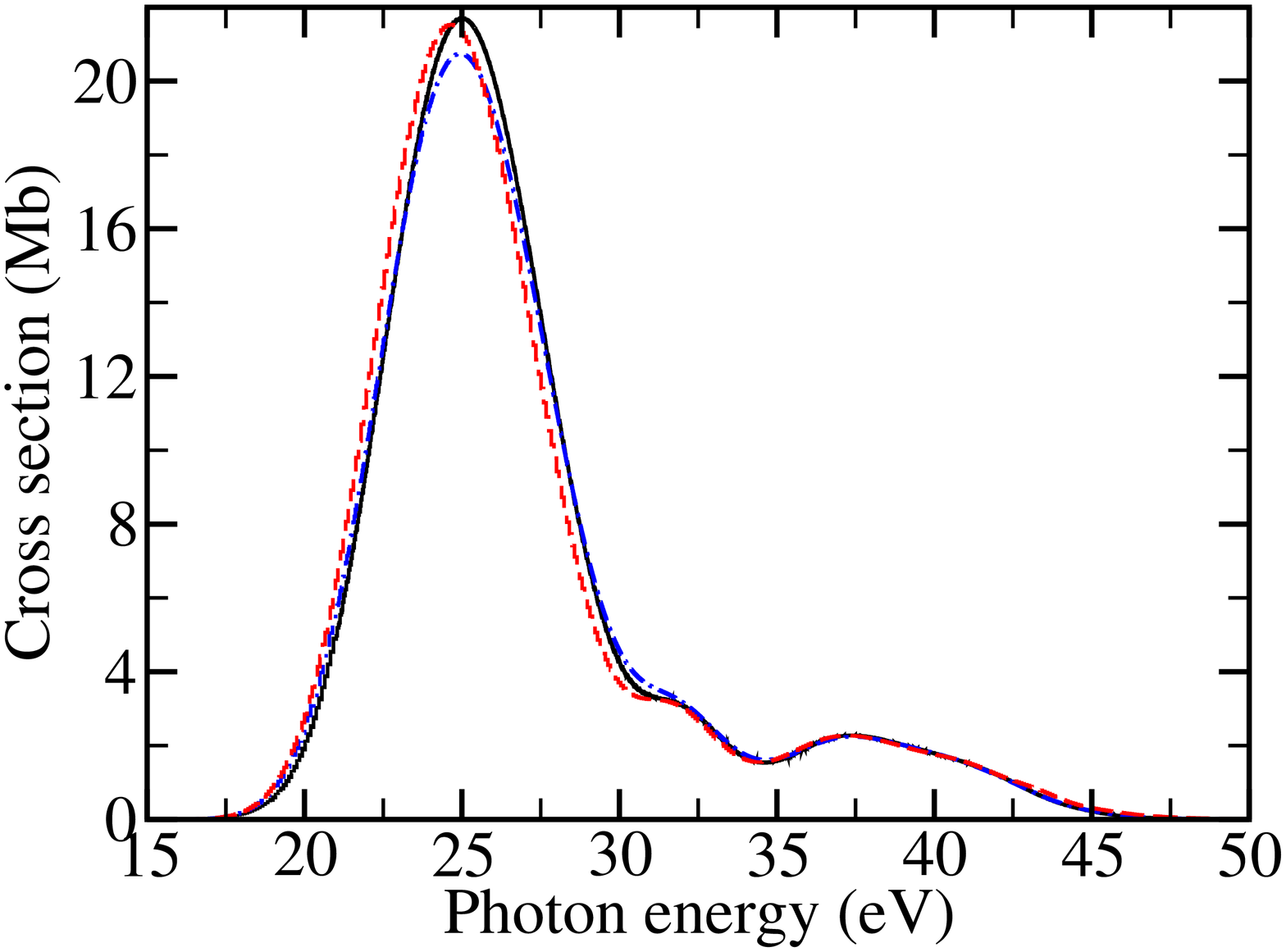}}}
\caption{Comparison of the results obtained for the parallel orientation of 
the molecular axis with respect to the field to the results from 
\cite{dia:saen03}:
the full black line is the result of the present calculation for $J = 1$ 
$\rightarrow$ $J' = 0, 2$, the blue chain is present calculation for $J = 0$ 
$\rightarrow$ $J' = 1$, and the red dashed line is the result from  
\cite{dia:saen03} (obtained for  $J = 0$ $\rightarrow$ $J' = 1$).} 
\label{comp}
\end{figure}

In figure \ref{comp} the calculated photodissociation cross section for a 
parallel alignment of the molecular axis with respect to the photon 
polarization vector ($g=1$) is compared to a previous result \cite{dia:saen03} 
in which explicitly correlated basis functions had been used for the
evaluation of the electronic wavefunctions. Both results were obtained 
for transitions starting from the absolute rovibronic ground state 
of the HeH$^+$ molecule ($\nu=0$, $J=0$). The most evident difference 
between the two spectra lies in the position and height of their maxima. 
This difference is due to the fact that the electronic
wavefunctions especially for the ground and the first excited states are 
more accurately represented in \cite{dia:saen03} than in the present 
calculation, as is concluded from a comparison of the corresponding 
energies. 

In references\,\cite{dia:vann04,dia:vann06} it was demonstrated that 
the present approach can yield extremely accurate two-electron wavefunctions, 
but this requires a judicious choice of the basis set, if the size of the 
calculation should be kept reasonably small. Most importantly, different 
electronic states possess different optimal basis sets. On the other hand,   
the goal of the present work was to consider as many excited bound states as
possible and to calculate all of them with sufficient and comparable 
accuracy. The most convenient way to do this is to use a single basis set 
for all states, 
since it avoids non-orthogonality problems and preserves sum-rule 
completeness. This leads to a compromise for the results for the ground and 
first excited states. It was checked that by improving 
the ground and first excited states the agreement to \cite{dia:saen03} 
in the peak region is also improving. The remaining part of the spectrum that
stems from transitions to higher-lying electronic states is, however, 
in good agreement with the previous result in \cite{dia:saen03}. As is
discussed below, for a comparison to the experiment reported in
\cite{dia:pede07} the accuracy of the first excited electronic state plays 
practically no role and thus the good agreement with the previous calculation for 
the higher photon energies is more important for the purpose of this work.  

Figure \ref{comp} shows also the parallel photodissociation cross section for the 
first excited rotational state ($J=1$) which, according to \cite{dia:basu84}, 
is the most probable rotational level at room temperature. Accidentally, the 
result for $J=1$ agrees better to the previous theoretical calculation 
performed for $J=0$. From this accidental agreement and the discussion above 
it is clear that the influence of rotational excitation is in general rather 
small and most prominent for the transition to the first excited electronic 
state. It was, however, checked that the results shown in the following are 
all relatively insensitive to the initial rotational state, as long as $J$ is 
not very large.

\begin{figure}[th]
\epsfxsize 1.00\hsize
\centerline{\mbox{\epsfbox{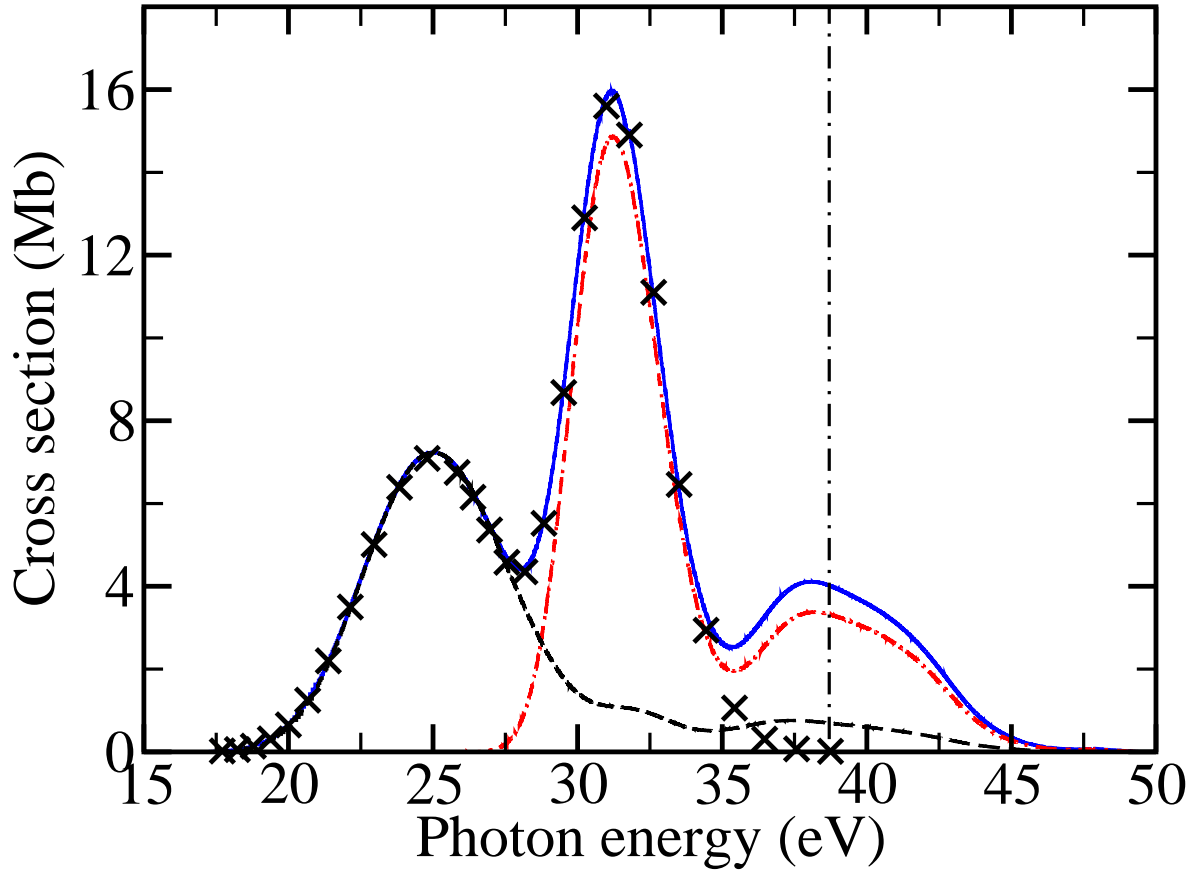}}}
\caption{Total photodissociation cross section of HeH$^{\bm +}$
         after averaging over an 
         isotropic orientational distribution (blue solid). 
         Also shown 
         are the partial contributions from parallel (black dashes) and 
         perpendicular (red chain) orientations. 
         The calculation is performed 
                for transitions starting from the lowest lying vibrational 
                state $\nu = 0$ and from $J = 1$.The crosses are the
         theoretical result from \cite{dia:basu84} and the vertical chain 
         at 38.7\,eV marks the experimental photon energy.} 
\label{iso}
\end{figure}

Figure \ref{iso} shows the photodissociation cross section of HeH$^{\bm +}$ 
for an isotropic orientation of the molecular axis with respect to the field, 
as well as the parallel ($g=1/3$) and perpendicular ($g=2/3$) contributions 
to it, for transitions starting 
from the lowest lying vibrational state $\nu = 0$ and the first excited 
rotational level $J = 1$. It can be seen in figure \ref{iso} that 
until about 27.5\,eV the spectrum is practically identical to the parallel 
contribution, while above about 29\,eV the perpendicular contribution 
is clearly dominant. Similarly to the parallel case the perpendicular 
spectrum shows a dominant low-energy peak followed by a rather structureless 
tail. In fact, the tail looks more like a second peak, but it is composed of 
a number of electronic transitions. In comparison to the tail for a parallel 
alignment the one in the perpendicular case is confined to a smaller energy 
window. 

An older calculation \cite{dia:basu84} (performed using the length gauge)
took into account only three excited
states of HeH$^{\bm +}$, namely the first two excited $\Sigma$ states and the
first $\Pi$ state. Figure \ref{iso} shows that for these particular states
the agreement between the present calculation and the one from 
\cite{dia:basu84} is excellent, but at the photon energy used in the FLASH
experiment higher lying states clearly dominate the spectrum.

\subsection{Comparison to experiment: channel separation, diabatic effects, 
and vibrational excitation}
The heteronuclear HeH$^{\bm +}$ ion can dissociate into two different channels:  
He + H$^{\bm +}$ and H + He$^{\bm +}$.
The FLASH experiment \cite{dia:pede07} measured only the cross 
section given by the dissociation of HeH$^{\bm +}$ into He + H$^{\bm +}$. 
A comparison requires thus a separation of the two channels. 
Within the adiabatic approximation the asymptotic long-range behaviour, 
i.\,e.\ the separated atom limit at $R\rightarrow\infty$, defines the 
channel into which the population of a given adiabatic electronic state 
dissociates. 
At $R=60\,a_0$ the potential curves were found to have almost reached the 
separated atom limit. A comparison of the asymptotic potential-curve energies 
with the sum of the energies of an independent atom and ion (He + H$^{\bm +}$ or 
H + He$^{\bm +}$) in their various states yielded a well-defined channel 
assignment. The result is shown in figures~\ref{potsig} and \ref{potpi}. 
Most of the lower lying states dissociate into He + H$^{\bm +}$ while the 
highest lying Rydberg states dissociate into H + He$^{\bm +}$. The two 
blocks are rather well separated, except for three states (for both $^1\Sigma$ 
and $^1\Pi$ symmetry) that dissociate into H + He$^{\bm +}$ but lie in the 
energy regime dominated by the He + H$^{\bm +}$ channel.

\begin{figure}[th]
\epsfxsize 1.0\hsize
\centerline{\mbox{\epsfbox{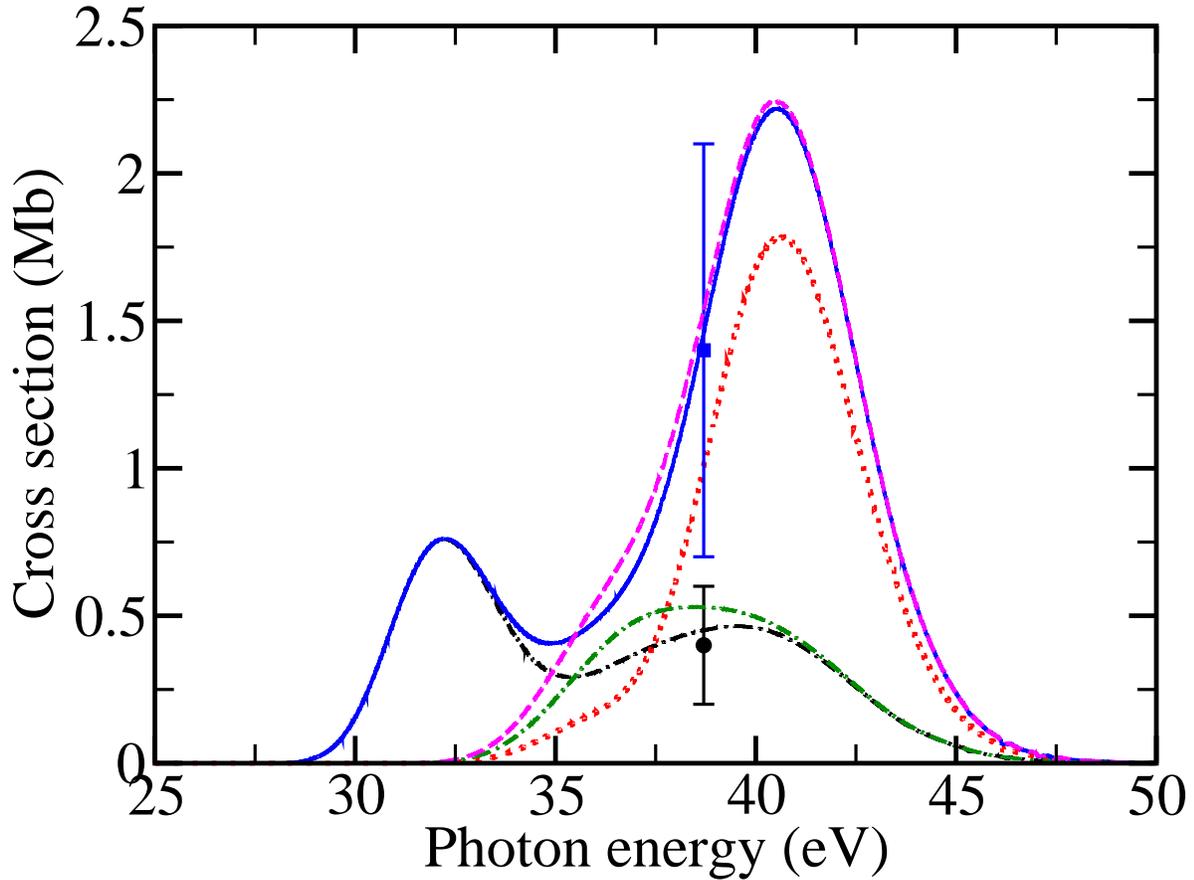}}}
\caption{Partial photodissociation cross section of HeH$^{\bm +}$ yielding  
                He {\bf +} H$^{\bf +}$. The calculations are performed 
                for transitions starting from the lowest lying vibrational 
                state $\nu = 0$ and from $J = 1$. The blue square is the 
                experimental total cross section and 
                the black circle is the experimental parallel contribution 
                (both from \cite{dia:pede07}). {\bf Adiabatic limit}: 
                the blue solid line 
                represents the isotropic orientation of the molecular axis 
                with respect to the field. The parallel (black chain) 
                and perpendicular (red dots) contributions are also shown. 
                {\bf Diabatic limit}: the magenta dashes represent the
                isotropic 
                orientation, while the green chain is the parallel
                contribution. 
                }
\label{chan1}
\end{figure}

After considering only transitions to electronic states that are assigned to 
the He + H$^{\bm +}$ channel the {\it partial} photodissociation cross section 
shown in figure \ref{chan1} is obtained. The result (obtained for 
$\nu=0$ and $J=1$) shows in comparison to figure \ref{iso} that overall 
dissociation into the He + H$^{\bm +}$ channel is clearly less favoured. 
However, the relative importance of the two channels varies strongly with 
photon energy. Below about 27\,eV the spectrum stems solely from the 
first excited state of $^1\Sigma$ symmetry which dissociates into 
H + He$^{\bm +}$. In between about 30 and 35\,eV the total dissociation 
cross section is dominated by the contribution from the lowest lying 
$^1\Pi$ state that also dissociates into H + He$^{\bm +}$. A smaller 
contribution comes, however, from the second excited $^1\Sigma$ state. 
This state is responsible for the first peak observed in the He + H$^{\bm +}$ 
channel in figure \ref{chan1}. Thus the parallel contribution is equal to 
the total cross section of this channel in that energy range. Above about 
35\,eV many other states contribute, and the cross sections for the two 
dissociation channels are almost of equal size, as 
a comparison of figures \ref{iso} and \ref{chan1} shows. Since the 
majority of the states dissociating into He + H$^{\bm +}$ lies energetically 
above the ones dissociating into H + He$^{\bm +}$, the maximum of the 
second peak of the He + H$^{\bm +}$ cross section is shifted to slightly 
higher energy than the one of the total dissociation cross section.     

For comparison, figure \ref{chan1} shows also the experimentally 
determined {\it absolute} partial dissociation cross section at 38.7\,eV 
for the He + H$^{\bm +}$ channel as well as its parallel contribution
\cite{dia:pede07}. The agreement is almost perfect, but further aspects 
have to be considered for a direct comparison. First, the present analysis 
is based on an adiabatic model. Second, it appears to be clear that the 
HeH$^+$ ions investigated in the experiment were not in their vibrational 
ground state. These two issues are thus discussed in the remainder of this 
section.   

The potential curves in figures \ref{potsig} and \ref{potpi} show 
a number of avoided crossings. Clearly, at these crossings there is some 
probability for a transition from one adiabatic curve to the other. In the 
present context not all avoided crossings are, however, of equal importance. 
Those between states dissociating into the same dissociation channel do 
not change the partial cross section into, e.\,g., the He + H$^{\bm +}$ 
channel. (Clearly, these avoided crossing have to be considered in a 
calculation of other differential spectra like the kinetic-energy distribution 
of the ionic or atomic fragments.) A more careful analysis that takes also the 
relative importance of the different electronic states (at the experimental 
photon energy) into account shows that the most critical avoided crossing for 
the present discussion is the one occuring close to $R=4.0\,a_0$ between the 
2nd and 3rd excited state of $^1\Sigma$ symmetry (see figure \ref{potsig}).  
(This avoided crossing is already visible in the theoretical data in 
\cite{dia:gree74a,nu:jons99} and thus reproduced by other numerical 
methods.) 

A standard procedure for estimating the importance of diabatic effects at an 
avoided curve crossing is provided by the Landau-Zener approximation (see,
e.\,g., Chapter 10 in \cite{gen:bill96} and references therein). Within 
this model the probability for a transition from one adiabatic curve to the
other is given by 
\begin{equation}
\label{eq:zener}
P_{12}^{a} = \exp \left[-\,\frac{2 \pi H_{12}(R_{c})^{2}}{\hbar v 
             |F_1 - F_2|} \right]\;,
\end{equation} 
where $R_{c}$ is the internuclear separation at which the avoided crossing 
occurs, $H_{12}(R_{c})$ is the diabatic coupling which can be calculated from 
the relation $E_2 - E_1 = 2 H_{12}$ ($E_1$ and $E_2$ being the adiabatic 
energies), the velocity $v$ is 
obtained from the kinetic energy and $F_{1,2}$ are the slopes 
of the diabatic curves at $R_{c}$. 
In the present case, it turns out that 
the Landau-Zener formula is difficult to apply, since the two curves 
cross in a region where their behaviour is not 
at all linear. (A removal of the nuclear repulsion term does not resolve 
this problem.) As a consequence, it is difficult to estimate the slopes 
and the result of the Landau-Zener estimate has such a large error bar 
that it is practically needless.  
Consequently, figure \ref{chan1} shows the cross sections in the two extreme 
cases: the already discussed adiabatic limit ($P_{12}^{a}=0$) and the fully 
diabatic one ($P_{12}^{a}=1$). From the figure one may conclude 
that the influence of possible diabatic effects on the cross section at 
the photon energy used in the experiment is not very relevant. The most 
noticeable changes would take place at lower photon energies. 

\begin{figure}[t]
\epsfxsize 1.00\hsize
\centerline{\mbox{\epsfbox{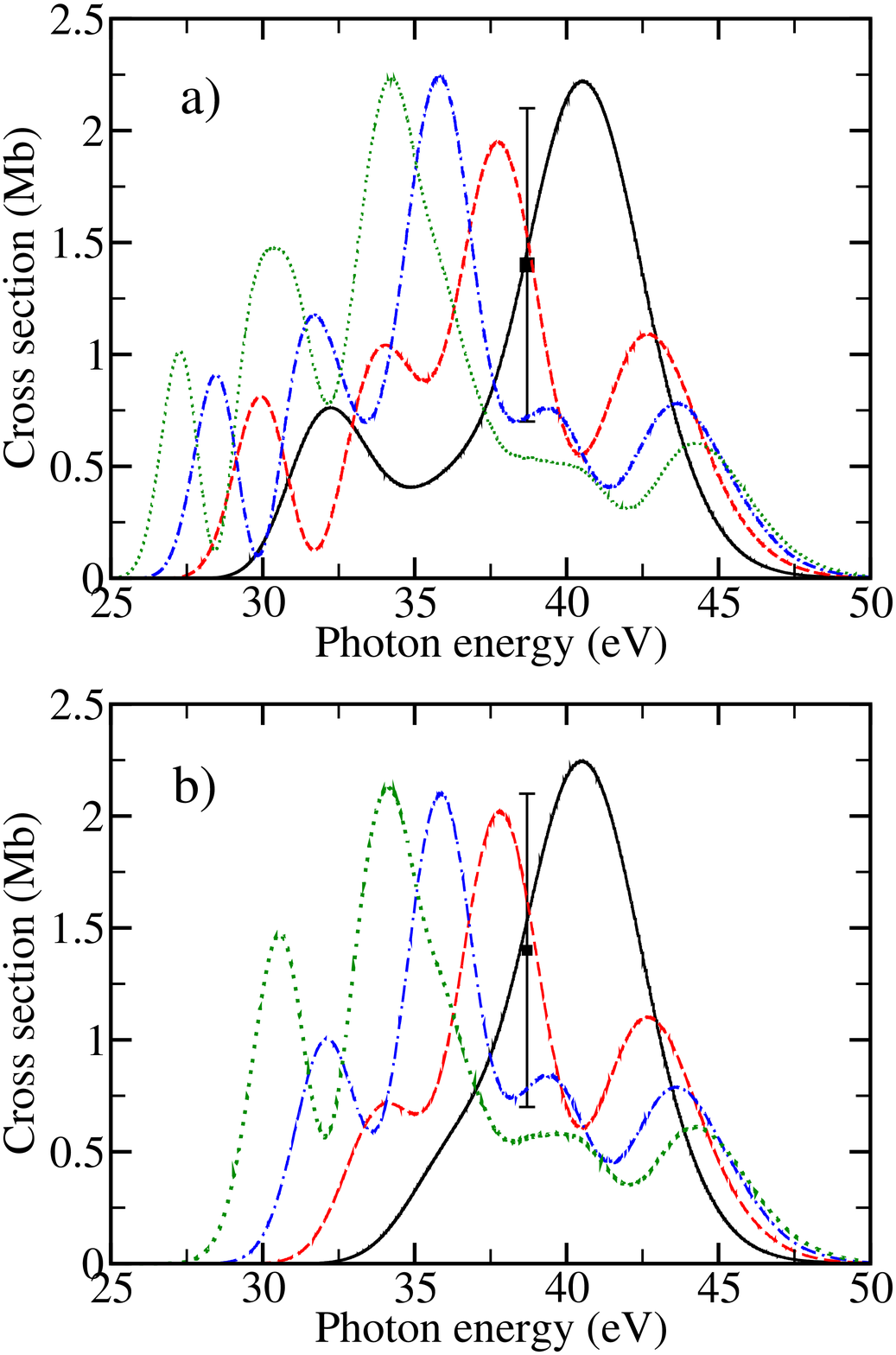}}}
\caption{Partial photodissociation cross section of HeH$^{\bm +}$ decaying 
                into He {\bf +} H$^{\bf +}$ in a) the adiabatic and b) the 
                diabatic limit for an isotropic orientation and different 
                initial vibrational states: $\nu = 0$ 
                (full black line), $\nu = 1$ (red dashes), 
                $\nu = 2$ (blue chain), and $\nu = 3$ (green dots). 
                The black square is the experimental result \cite{dia:pede07}.}
\label{n1n4}
\end{figure}

All the spectra shown so far were obtained by assuming that the system 
starts in the lowest lying vibrational state $\nu = 0$. 
This is valid for thermal distributions at not too high temperatures. 
However, due to the way the HeH$^+$ ions are generated in the FLASH 
experiment reported in \cite{dia:pede07} the detailed vibrational 
distribution is unknown, but not expected to be thermal. In a more 
recent experiment on He$_2^+$ in which the ions were produced in the 
same way (by heavy particle collisions) the vibrational distribution 
was determined \cite{dia:buhr08}. In this case the initial-state 
distribution was found to be distributed over at least 5 vibrational 
states ($\nu=0$ to 4). Since the binding energy of He$_2^+$ 
is similar to that of HeH$^+$, one may expect a similar 
vibrational-state distribution for both ions. Therefore, photodissociation 
cross sections for transitions starting from different vibrational states 
were computed. The ones corresponding to the initial states $\nu = 0,1,2$, 
and 3 are shown in figure \ref{n1n4}.

At the experimental photon 
energy the results for both the adiabatic and diabatic limit are very 
close to each other, but there are evident differences at lower photon 
energies. Accidentally, the photon energy in \cite{dia:pede07} probes 
the spectrum of the HeH$^+$ ion at a point where the cross sections 
of the $\nu=0$ and $\nu=1$ states cross each other. As a consequence, 
the performed experiment is rather insensitive to a redistribution 
of population between the states with $\nu=0$ and $\nu=1$.   
The population of higher vibrational states ($\nu>1$) leads on the 
other hand to a decrease of the cross section at 38.7\,eV. The trend 
visible for $\nu=2$ and $\nu=3$ continues for higher values of 
$\nu$, as was checked though not explicitly shown. 

\clearpage

\begin{figure}[th]
\epsfxsize 1.00\hsize
\centerline{\mbox{\epsfbox{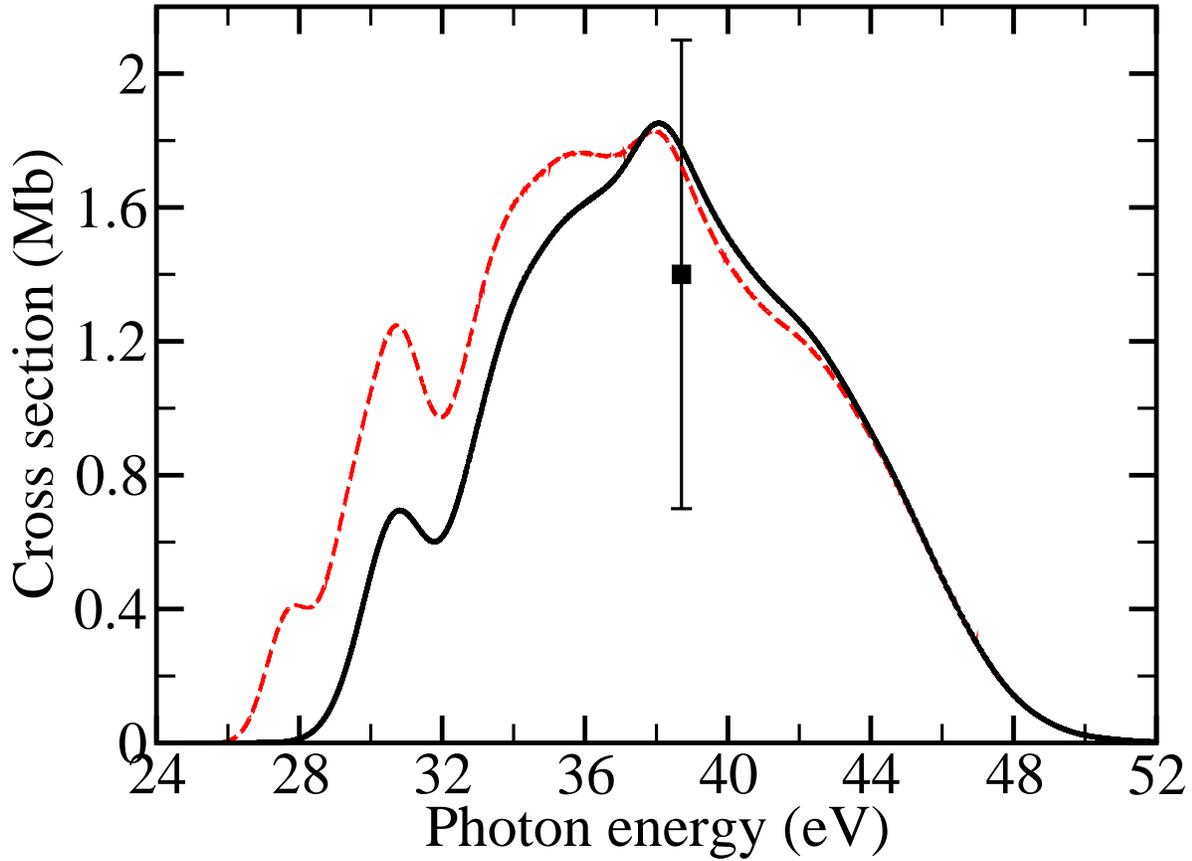}}}
\caption{Photodissociation of HeH$^{\bm +}$ into 
         He {\bf +} H$^{\bf +}$ - comparison between theoretical calculations 
         (adiabatic - red dashes, and diabatic estimate - 
         black solid line) for transitions starting from a mixture 
         of vibrational levels (33$\%$ $\nu = 0$, 45$\%$ $\nu = 1$, 
         and 22$\%$ $\nu \geq 2$) and initial rotational level $J = 1$ 
         with the experimental (black square) result \cite{dia:pede07}.} 
\label{mixvib}
\end{figure}

If one assumes the initial vibrational distribution of HeH$^+$ in 
\cite{dia:pede07} to be identical to the one of He$_2^{2+}$ in 
\cite{dia:buhr08}, the photodissociation cross section shown 
in figure \ref{mixvib} is obtained. The vibrational distribution 
is in this case modelled in the following way: 
33$\%$ of the vibrational population was considered to be in the $\nu = 0$ 
level, 45$\%$ in $\nu = 1$, and 22 $\%$ are equally distributed over 
the states $\nu = 2,3,4$. Compared to figure \ref{chan1} the influence 
of the vibrational distribution is obvious: the shape of the spectrum  
changes and the value of the cross section at the experimental photon energy 
(38.7\,eV) increases slightly (from $\sigma \approx 1.50$ Mb -- see figure 
\ref{chan1} -- to $\sigma \approx 1.70$ Mb). Thus the agreement to experiment 
decreases, but remains still well within the error bars of the measured 
cross section. In view of the lack of knowledge about the experimental 
initial vibrational distribution, it is, however, impossible to finally 
conclude on the degree of agreement between theory and experiment. 
As the present 
study shows, this distribution has a rather strong influence on the 
photodissociation cross section. In fact, the theoretical results 
may indicate a narrower vibrational distribution than the one 
found for He$_2^{2+}$ in \cite{dia:buhr08} that was assumed to be 
similar to the one of HeH$^+$ in \cite{dia:pede07}. It would certainly 
be of interest to have experimental data also for other photon energies. 
This may also help to (indirectly) determine the initial vibrational-state 
distribution. Furthermore, these data may be an interesting direct measure 
of the adiabaticity at the discussed avoided crossing between the 2nd and 
3rd excited state of $^1\Sigma$ symmetry.


\section{Summary}
\label{summ}
In this work the photodissociation cross section of HeH$^{\bm +}$ was 
computed. A previous theoretical result that considered a parallel 
alignment of the molecular axis was confirmed. A calculation for 
a perpendicular alignment was performed that considered a large 
number of electronic states in order to obtain a converged photodissociation 
cross section. In order to compare to a recent experiment, the partial 
cross sections for dissociation into either He + H$^+$ or H + He$^+$ 
were determined. This was done within the adiabatic approximation, but 
possible diabatic effects were estimated. 
Different initial vibrational and rotational levels were considered 
and photodissociation spectra for 
transitions starting from the rotational levels $J = 0,1$ and vibrational 
levels $\nu = 0,1,2,3,4$ were presented. The influence of the vibrational 
motion on the final spectra was investigated and compared to the 
experimental findings. Agreement within the experimental error bar 
was found, but it was concluded that the knowledge of the experimental 
vibrational distribution is important for a final conclusion. The 
present analysis should motivate further experimental studies, 
especially at variable photon energy, as it should become possible 
due to the current progress in the development of free-electron 
lasers.


\section*{Acknowledgments}
The authors acknowledge helpful discussions with A.~Wolf and financial support 
by the {\it Deutsche Forschungsgemeinschaft} through the SFB 450, 
the {\it Stifterverband f\"ur die Deutsche Wissenschaft}, the 
European COST Action CM0702, and the {\it Fonds der Chemischen Industrie}.


\section*{References}


\end{document}